\documentclass[aps,prl,twocolumn,superscriptaddress,showpacs]{revtex4-1}
\usepackage[pdftex]{graphicx}
\usepackage{graphicx}
\usepackage{dcolumn}
\usepackage{bm}
\usepackage{amssymb}
\makeatletter
\def\captionof#1#2{{\def\@captype{#1}#2}}
\makeatother

\usepackage[english]{babel}
\usepackage{times}
\usepackage{graphicx}
\usepackage{amsfonts}
\usepackage{psfrag}
\usepackage{verbatim}
\usepackage{color}

\begin{document}

\title{Thermal phase transitions  in Artificial Spin-Ice}
\author{Demian Levis} 
\affiliation{Universit\'e Pierre et Marie Curie
  - Paris 6, Laboratoire de Physique Th\'eorique et Hautes Energies,
  4, Place Jussieu, Tour 13, 5\`eme \'etage, 75252 Paris Cedex 05,
  France}   
 
 \author{Leticia F. Cugliandolo} 
\affiliation{Universit\'e Pierre et Marie Curie
  - Paris 6, Laboratoire de Physique Th\'eorique et Hautes Energies,
  4, Place Jussieu, Tour 13, 5\`eme \'etage, 75252 Paris Cedex 05,  France}
  
 \author{Laura Foini} 
\affiliation{Universit\'e Pierre et Marie Curie
  - Paris 6, Laboratoire de Physique Th\'eorique et Hautes Energies,
  4, Place Jussieu, Tour 13, 5\`eme \'etage, 75252 Paris Cedex 05,  France}

 \author{Marco Tarzia} 
\affiliation{Universit\'e Pierre et Marie Curie
  - Paris 6, Laboratoire de Physique Th\'eorique de la Mati\`ere Condens\'ee,
  4, Place Jussieu, Tour 12, 5\`eme \'etage, 75252 Paris Cedex 05,  France}


\begin{abstract}
We use the sixteen vertex model to describe 
bi-dimensional artificial spin ice (ASI). We find excellent
agreement between vertex densities in fifteen 
differently grown samples and the predictions of the model. Our results demonstrate that
the samples are in usual thermal equilibrium away from a 
critical point separating a disordered and an anti-ferromagnetic phase
in the model. The second-order phase transition that we predict 
suggests that the spatial arrangement of 
vertices in near-critical ASI should be studied in more detail 
in order to verify whether they show the expected space
and time long-range correlations.
\end{abstract}

\pacs{75.50.Lk, 75.10.Hk, 05.70.Jk, 05.70.Ln}
\keywords{Suggested keywords}

\maketitle

\setlength{\textfloatsep}{10pt} 
\setlength{\intextsep}{10pt}

 
Hard local constraints produce a rich variety of collective behavior such as the 
splitting of phase space into different topological sectors and  the existence of 
``topological phases'' that cannot be described with conventional order parameters~\cite{Balents2010}.
In geometrically constrained magnets, the local minimization of the interaction energy on a frustrated  unit 
gives rise to a macroscopic degeneracy of the ground state~\cite{MoessnerRamirez}, 
unconventional phase transitions~\cite{Jaubert2008-short, LiebWuBook}, 
the emergence of a ``Coulomb" phase with long-range correlations~\cite{Youngblood1980-short,Henley2010} 
and slow dynamics~\cite{Fennell2005-short, Levis2012}
 in both $2D$ and $3D$ systems.
In this work we focus on a paradigm with these features: 
\emph{spin-ice}, a class of magnets frustrated  by the 
\emph{ice-rules}~\cite{Harris1997-short, Bramwell2001d,DiepBookCH7}.

In {\it natural spin-ice}  the  ice rules  are due to two facts: the 
pyrochlore lattice structure, in which rare earth magnetic ions sit on the vertices of 
corner sharing tetrahedra, and the ferromagnetic interaction between the Ising-like moments
that are forced to lie on the edges joining the centers of neighboring tetrahedra
by crystal fields.  The energy of each unit cell 
is thus minimized by configurations with two spins pointing 
in and two out of the center of the cell. All configurations 
satisfying the local constraint are degenerate ground states if the ice-rule preserving vertices are equally probable. 
This is the same mechanism whereby water ice has a non-vanishing zero-point entropy that is, indeed,   
remarkably close to the one of the  Dy$_2$Ti$_2$O$_7$ spin-ice compound~\cite{Ramirez1999-short}. 
These materials received a renewed interest in recent years
when a formal mapping to magnetostatics suggested to interpret the 
local configurations violating the spin-ice rule as magnetic charges~\cite{Castelnovo2008a-short}. 
However, the detailed study of such defects in $3D$ 
remains a hard experimental task.
 
Bi-dimensional Ising-like ice-models had no experimental counterpart until 
recently when it became possible to manufacture artificial samples made of arrays of 
elongated single-domain ferromagnetic nano-islands frustrated by 
dipolar interactions.
The beauty of {\it artificial spin-ice} (ASI) is that the 
interaction parameters can be precisely engineered---by tuning the distance between islands, 
{\it i.e.}~the lattice constant, or by applying external fields---and 
the microstates can be directly imaged with magnetic force 
microscopy (MFM)~\cite{Wang2006-short}. These systems should set into 
different phases depending on the experimental conditions~\cite{Moller2006}.
However, the lack of thermal fluctuations due to the high energy barriers for
single spin-flips, had prevented the observation of the expected two-fold degenerate antiferromagnetic (AF) 
ground state. Recently, 
these problems have been partially overcome {\it via} 
(i) the gradual magneto-fluidization of an initially polarized state~\cite{Nisoli2010a-short} and   
(ii)  the thermalization of the system during the slow growth of the 
samples~\cite{Morgan2011-short}. Although with de-magnetization  
the actual AF state of square ASI was never reached, the statistical 
study of a large number of frozen configurations of samples with up to $10^6$ vertices 
at interaction-dominated low energies 
became possible~\cite{Nisoli2010a-short}.
With the procedure proposed in (ii) 
large regions with  AF order were formed in a few samples when sufficiently small lattice constant and 
weak disorder were used. Whether the sampling of a conventional thermal equilibrium ensemble is achieved in this way is 
a question that was raised in~\cite{Morgan2011-short,Morgan2013} and that we will address here. 

\begin{figure}[t]
\begin{centering}
\includegraphics[scale=1]{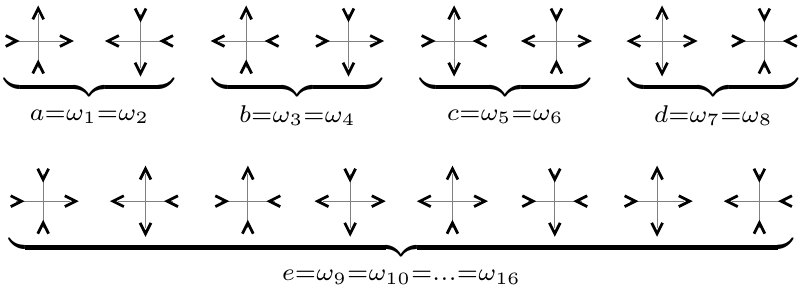}
\end{centering}
\caption{
Sixteen  vertex configurations and their statistical 
weights $\omega_i \propto \exp{(-\beta \epsilon_i)}$. The first six-vertices, $v_1,\dots v_6$, 
verify the ice-rule and have vanishing magnetic charge. Vertices $v_1, \dots, v_4$ carry a dipolar moment while 
$v_5$ and $v_6$ do not. The last ten vertices are ``defects'': $v_7$ and $v_8$ have magnetic charge 
$\pm 4$ and no dipole moment while  $v_9, \dots, v_{16}$ have magnetic charge $\pm 2$ and a net dipole moment.}
\label{16vertex}
\end{figure}

The purpose of this letter is to interpret and explain very recent experimental observations on 
ASI~\cite{Nisoli2010a-short,Morgan2011-short,Morgan2013}.
With this aim we consider a simple schematic model for $2D$ dipolar spin-ice,
the \emph{sixteen-vertex model} (see Fig.~\ref{16vertex})~\cite{LiebWuBook,Levis2012}, where dipolar interactions
beyond nearest-neighbor vertices are neglected.
We compare the equilibrium and out-of-equilibrium properties of the model with the experimental 
results and
we match the behavior of several observables (such as the densities of different vertex types
$\langle n_i \rangle$) with data on ASI samples~\cite{Morgan2011-short,Morgan2013}.
This approach is at face value similar but actually very different from the one used
in previous studies~\cite{Nisoli2010a-short,Morgan2013}, where
data were fitted in terms
of {\it single independent vertices} (with no topological interactions), 
as $\langle n_i \rangle \propto \exp(-\beta_{\rm eff} \epsilon_i)$,
with the effective temperature $\beta_{\rm eff}$ introduced as a fitting parameter.
On the contrary, in our approach frustrated interactions are fully
taken into account.
This analysis allows us to address the issue of thermalization of ASI samples and
to make several predictions
that could be tested in the lab. 

\begin{figure}[t]
\begin{centering}
\includegraphics[scale=0.8]{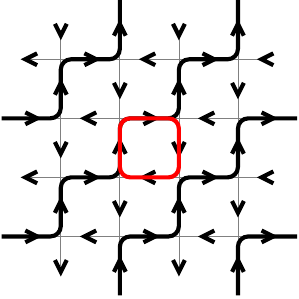}(a)\ \ 
\includegraphics[scale=0.8]{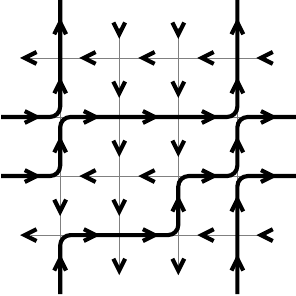}(b)\ \ 
\includegraphics[scale=0.8]{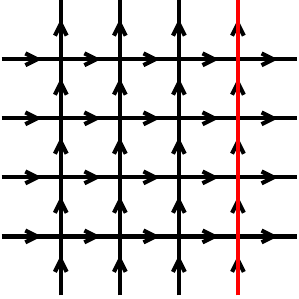}(c)
\end{centering}
\caption{(Color online.)
Characteristic configurations of the ground state ($d=e=0$) of three different phases of the 
sixteen-vertex model on the square lattice. (a)~Antiferromagnetic ($c$-AF) order. Reversing the 
central (red) loop yields an elementary excitation. (b)~Collective spin liquid (SL) phase. (c)~Frozen 
ferromagnetic (FM) order. The reversed vertical string (red line) is an extended excitation.}
\label{fig:configurations}
\end{figure}

\paragraph{The samples and the model.} 
In their simplest setting ASI are $2D$ arrays
of elongated single-domain permalloy islands whose
shape anisotropy defines Ising-like spins arranged along
the edges of a regular square lattice. Spins interact through
dipolar exchanges and the dominant contributions are the ones between
neighboring islands across a given vertex.
The sample is frustrated since no configuration of the surrounding spins can
minimize all pair-wise dipole-dipole interactions on a vertex.
In samples with no height offset ($h=0$) 
$2D$ square symmetry defines four relevant vertex types
of increasing energy,
where the $c$ vertices (see Fig.~\ref{16vertex})
take the lowest value, leading to a ground state 
with staggered $c$-AF order [see Fig.~\ref{fig:configurations}(a)]. 
Note that the relative energies of the different vertex configurations could be tuned by $h$ in such 
a way that the ground state displayed FM order [see Fig.~\ref{fig:configurations}(c)]~\cite{Moller2006}. 

We  mimic the experimental samples with the sixteen-vertex model
defined as follows: Ising spins sit along the edges of an $L\times L$ square lattice. 
Long-range interactions beyond next nearest neighbor spins 
are neglected and the energies of the sixteen vertex configurations are attributed as 
explained above.
The total energy is given in terms of populations, $n_i$,
of distinct vertex types, $E=\sum_{i=1}^{16} n_i \epsilon_i$,
 each with a given magnetostatic energy $\epsilon_i$. 
Note that, despite the simple form of $E$, nontrivial frustrated 
topological interactions between vertices {\it are} present, due to the fact that 
each pair of neighboring vertices
shares the spin sitting along the edge joining them. 
The model is coupled to a thermal bath at inverse temperature $\beta$.
We introduce the statistical weights
$\omega_i \propto \exp(-\beta \epsilon_i)$, that in the literature of vertex 
models are usually referred as $a=$~$\omega_1=\omega_2$, 
$b$~$=$~$\omega_3=\omega_4$, $c=\omega_5=\omega_6$, $d=\omega_7=\omega_8$ and 
$e=\omega_9 = \dots = \omega_{16}$~\cite{LiebWuBook} (see Fig.~\ref{16vertex}).
The equilibrium properties of the model can thus be described in terms 
of four different statistical weights satisfying $c> (a=b)> e> d$.

We study the equilibrium and out-of-equilibrium properties of the sixteen-vertex model  
in two ways. We perform numerical simulations of the $2D$ model 
with a Monte-Carlo algorithm with {\it single-spin} updates improved with the rejection-free 
Continuous Time set-up (CTMC). 
We also employ an analytic approach put forward in~\cite{CavityMC-short}, 
based on a sophisticated version of a cluster variational 
mean-field Bethe-Peierls (CVBP) formalism, by defining the
model on a coordination-four tree made of square plaquettes (an extension over the 
tree of single vertices is necessary to correctly describe an AF phase
populated by  finite loop fluctuations, see Fig.~\ref{fig:configurations}). 
Details of the calculations have already been extensively reported in~\cite{CavityMC-short}, 
where the phase diagram was derived
for generic values of the ratios $a/c$, $b/c$, $d/c$, $e/c$, showing that
the CVBP technique describes with extremely good accuracy the equilibrium properties 
as compared to MC simulations.

\paragraph{Density of vertices.}
In the experiments in~\cite{Morgan2011-short,Morgan2013}
the thickness of the magnetic islands grows by deposition (at constant temperature
and all other external parameters within experimental accuracy) on fifteen lattices with five
different lattice constants and using three material underlayers (Si, Ti, Cr) in each of them. 
(We refer to the supplementary material in~\cite{Morgan2013}
for more details on the parameters and materials used.) The Ising
spins flip by thermal fluctuations during the growth process. However, as 
the energy barrier for single spin flips 
increases with the size of the islands, once a certain thickness is reached
the barrier crosses over the thermal fluctuations of the bath, $k_B T$,
and the spins freeze. At the end of the growth process, 
the  spin configurations are imaged with MFM and 
the number of vertices of each kind are counted 
on five independent square areas with, roughly, $27 \times 27$ vertices each. Average values
(and statistical errors) are also estimated.

\begin{figure}[t]
\begin{centering}
\includegraphics[scale=0.92,angle=0]{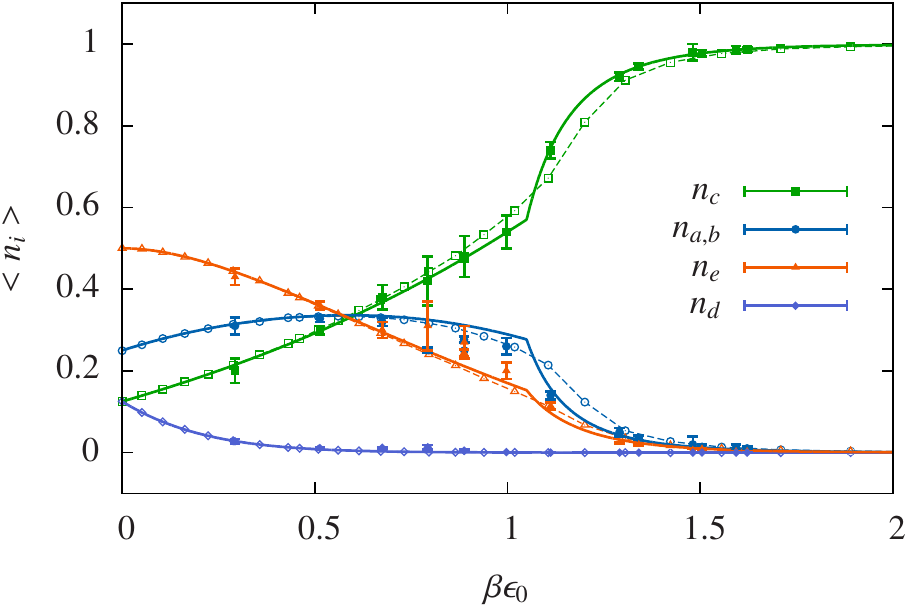}
\end{centering}
\caption{(Color online.)
Average densities of different vertex types as a function of $\beta \epsilon_0$. 
Full symbols with error bars are 
experimental data~\cite{Morgan2013}, $\langle n_i \rangle_{exp}$.
Empty symbols with dotted lines correspond to the equilibrium CTMC data, 
$\langle n_i \rangle_{sim}$. 
The CVBP analytic solution, $\langle n_i \rangle_{MF}$, 
of the sixteen-vertex model is shown in solid lines.}
\label{VertexData}
\end{figure}

Our analysis is as follows.
For each set of vertex concentrations measured in~\cite{Morgan2013} 
at a given lattice constant,
we determine the statistical weights $c$, $a$, $e$, and $d$---and thus 
 $\beta \epsilon_c$, $\beta \epsilon_a$, $\beta \epsilon_e$, and $\beta \epsilon_d$--- that better match the
data, by solving the sixteen-vertex model with the CVBP approximation
and imposing:
\begin{equation}
\langle n_i \rangle_{exp} = \langle n_i \rangle_{MF} = \frac{\sum_{\cal C} 
n_i e^{-{\cal H} (a,c,d,e)}}{\sum_{\cal C} 
e^{-{\cal H} (a,c,d,e)}} \, ,
\end{equation}
where the sum runs over all possible vertex configurations ${\cal C}$.
Surprisingly enough, it turns out that the energy ratios are approximatively constant within the
statistical errors for almost all
the lattice spacings of the as-grown samples, 
and coincide with the values used by Nisoli {\it et al.}~\cite{Nisoli2010a-short}.
Such energy ratios can be rationalized in term of magnetostatic exchanges due to dipolar interactions
between the islands of the sample associated to each vertex configuration. 
More precisely, each dipole is considered as a pair of oppositely charged monopoles sitting
on the vertices, and only Coulomb interactions between monopoles around a single vertex are taken into account,
yielding: $\epsilon_c = 2(1-2\sqrt{2})/\ell$, $\epsilon_a=-2/\ell$, $\epsilon_e = 0$,
$\epsilon_d = 2(2 \sqrt{2} + 1)/\ell$ ($\ell$ being the lattice constant). 
As a consequence, the statistical weights of the sixteen-vertex model
can be expressed in term of a {\it single energy scale} $\epsilon_0$ as
$c=1$, $a=e^{-\beta \epsilon_0}$, $e = e^{-\beta r_{e} \epsilon_0}$, and $d = e^{-\beta r_{d} \epsilon_0}$, 
with the energy ratios $r_e$ and $r_d$ being equal to
$r_e = (2 \sqrt{2} - 1)/(2 \sqrt{2} - 2) \simeq 2.207$ and $r_d = 2 \sqrt{2}/(\sqrt{2} - 1) 
\simeq 6.828$. 
Remarkably, no fitting parameter nor effective temperature needs to be introduced to describe the data: $\beta$ is the {\it true} inverse temperature
at which experiments are performed and $\epsilon_0$ is an energy scale that only depends on the lattice constant
and the underlayer, and which could be in principle determined from
a microscopic calculation.


\begin{figure}[t]
\begin{centering}
\includegraphics[scale=0.62]{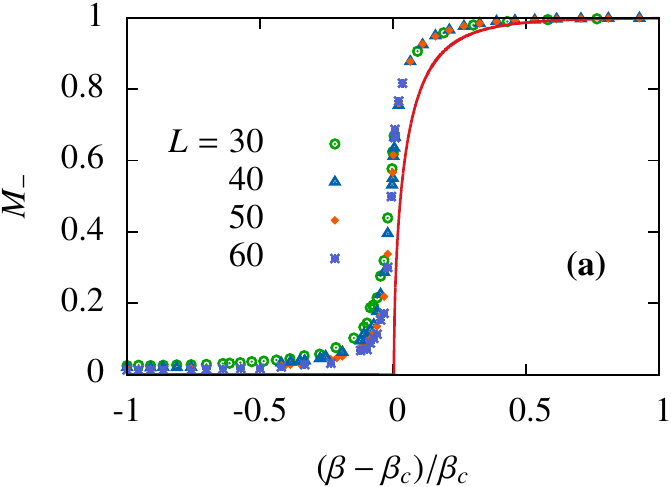}
\includegraphics[scale=0.62]{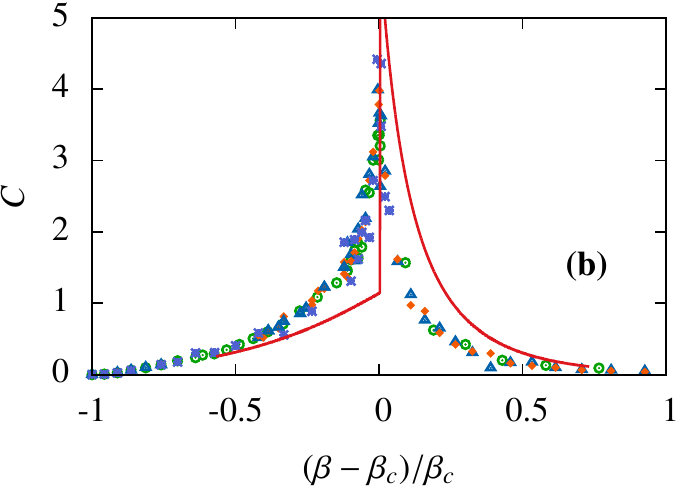}
\end{centering}
\caption{(Color online.) Staggered magnetization $M_-$~(a) and specific heat $C$~(b) as a function of the 
distance to the inverse critical temperature $\beta_c$ for system sizes 
$L=30, \ 40, \ 50, \ 60$. The (red) solid lines are the results of the CVBP analytic calculation. 
}
\label{FiniteSize}
\end{figure}

In Fig.~\ref{VertexData} we plot the thermal averages of the vertex densities 
 as a function of the energy scale $\beta \epsilon_0$. 
The CTMC results ($\langle n_i \rangle_{sim}$, empty symbols)
and CVBP calculations ($\langle n_i \rangle_{MF}$, solid lines) are in excellent agreement.
They show a second order phase transition from a paramagnetic (PM) phase at low
$\beta$ to an AF phase dominated by type-$c$ vertices at high $\beta$,
where both translational invariance and spin reversal symmetry are spontaneously broken.
We also include data from~\cite{Morgan2013} (full symbols) that are in
remarkable good agreement with the sixteen vertex model curves. 
(Except for a few points that turn out to be quite close to the critical temperature.) 

Such a good agreement confirms that dipolar interactions beyond nearest neighbor vertices 
do not play a prominent role and that the sixteen-vertex model mimics ASI samples. 
It also points out that the simple picture put forward in~\cite{Nisoli2010a-short} 
provides a good estimate of the energy ratios between different vertex types.
Finally, and most importantly, these results strongly suggest that the gradual growth of 
magnetic islands~\cite{Morgan2011-short,Morgan2013} seems to sample the conventional
thermal equilibrium ensemble for most of the experimental points.

\paragraph{Phases and phase transition.}
In the following we focus on the PM/AF phase transition.
In Fig.~\ref{FiniteSize} we  present equilibrium CTMC data for: (a)~the 
order parameter describing $c$-AF ordering, 
$M_-= \frac{1}{2} (\langle | m_{-}^{x}| \rangle + \langle | m_{-}^{y}| \rangle)$ where 
$m_-^{x,y}$ are the staggered magnetizations along the horizontal and vertical directions; (b)~the heat capacity 
$C=L^{-2}( \langle E^2\rangle -\langle E \rangle^2)$ as a function of 
the distance to the critical inverse temperature, $(\beta-\beta_c)/\beta_c$. 
The data clearly show the presence of a second-order phase transition 
from a conventional PM phase to a staggered AF phase as 
$\beta$ is increased above $\beta_c\epsilon_0$. 
The panels display the analytic results 
with solid red lines that yield a  
systematic shift of the critical point by about 10\% towards higher temperature, as expected
for mean-field calculations. We find $\beta_c^{MF} = 1.05$ and $\beta_c^{sim}= 1.2$, where 
here and in the following we measure $\beta$ in units of $\epsilon_0$.
 
 \begin{figure}[t]
\begin{centering}
\includegraphics[scale=0.68]{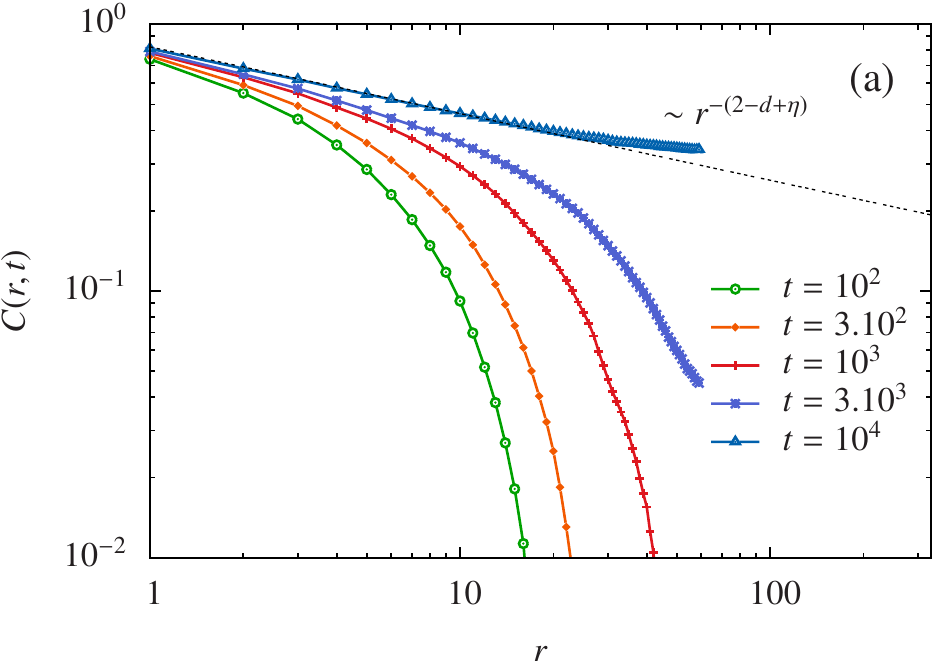}
\\
\hspace{0.25cm} \includegraphics[scale=0.35]{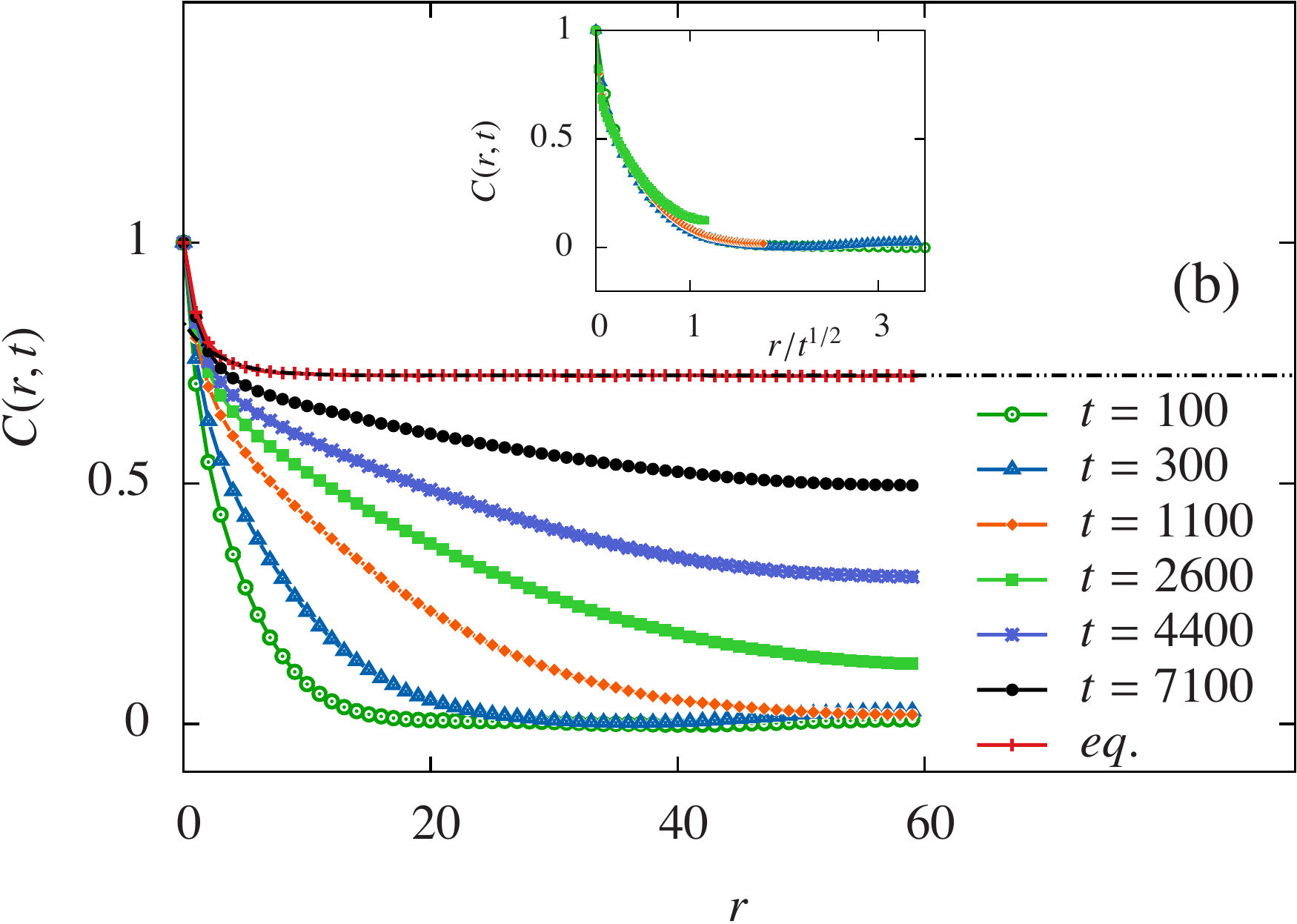}
\end{centering}
\caption{(Color online.)
Space-time correlations $C(r,t)$ after a quench from $\beta=0$ to 
 $\beta_c \approx 1.2$~(a) and 
$\beta=1.36$~(b), in a system with $L=60$. The colored lines-points are 
data taken at different times. The dotted black lines are the
$t \to \infty$ equilibrium functions, 
$C(r, t \to \infty) \sim r^{-2+d-\eta}$ with $\eta=1/4$,  
and $C(r, t \to \infty) \sim A \exp(-r/\xi)+M_-^2$, 
with $M_-^2\approx 0.73$, $\xi\approx 3$. The inset shows the data 
in panel~(b) where the $r$-axis has been rescaled by $t^{1/2}$.} 
\label{correlation-function}
\end{figure}

We determine $\beta_c=1.204\pm0.008$ independently
with a non-equilibrium relaxation method~\cite{Albano11-short}, by identifying 
$\beta_c$ as the inverse temperature at which  the staggered magnetization has an algebraic decay, 
$M_-(t)\sim t^{-\beta/(\nu z_c)}$ as a function of time, where $\beta$ and $\nu$ are the 
equilibrium critical exponents associated to the order parameter and the correlation length, respectively, 
and $z_c$ is the dynamical exponent. Away from this temperature an
exponential decay  is instead observed. This confirms
that the criticality of the SL phase~\cite{LiebWuBook, BaxterBook} 
is broken by the presence of defects at finite temperature.

 \begin{figure}[t]
\begin{centering}
\includegraphics[scale=0.22]{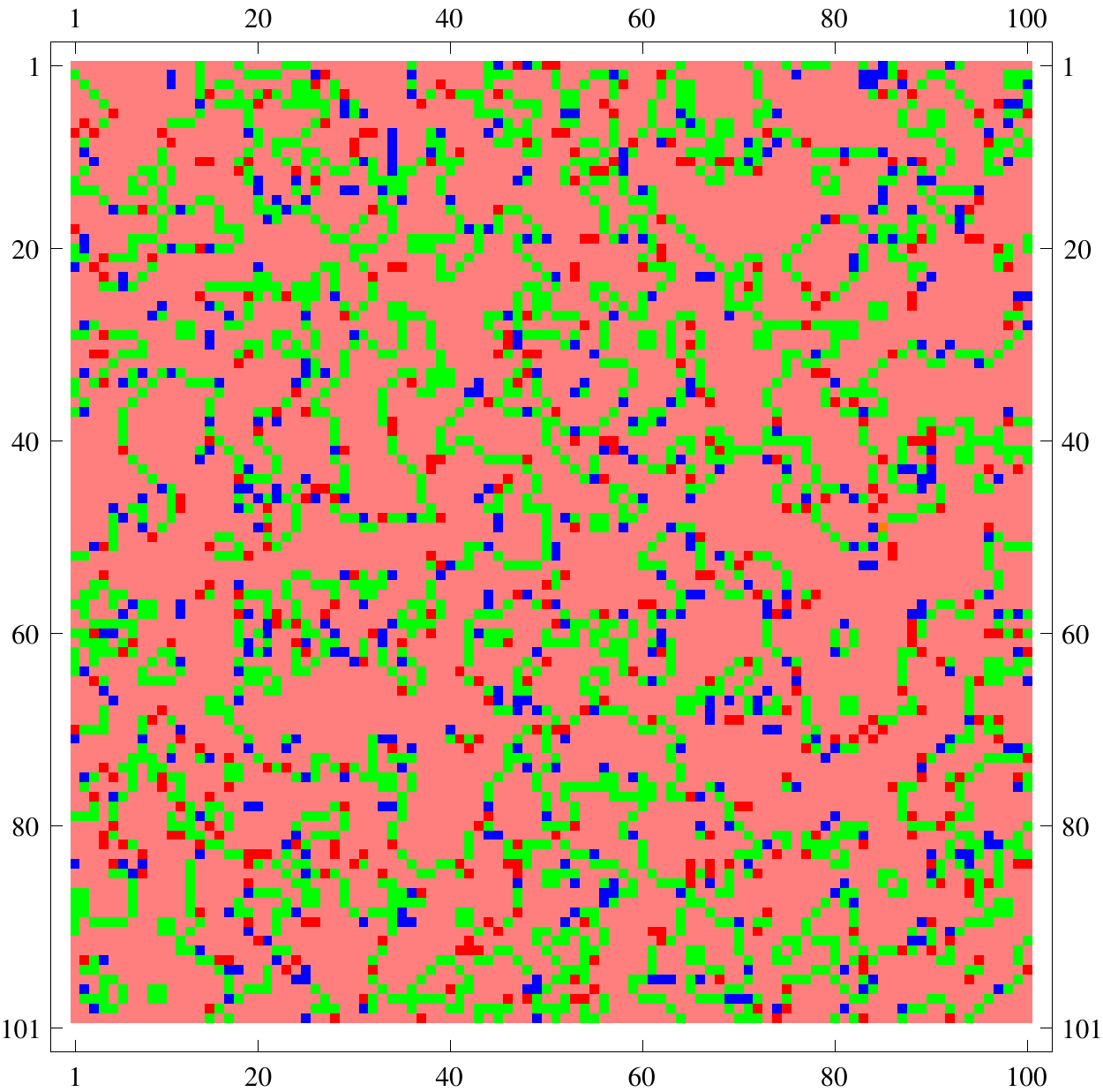}
\includegraphics[scale=0.22]{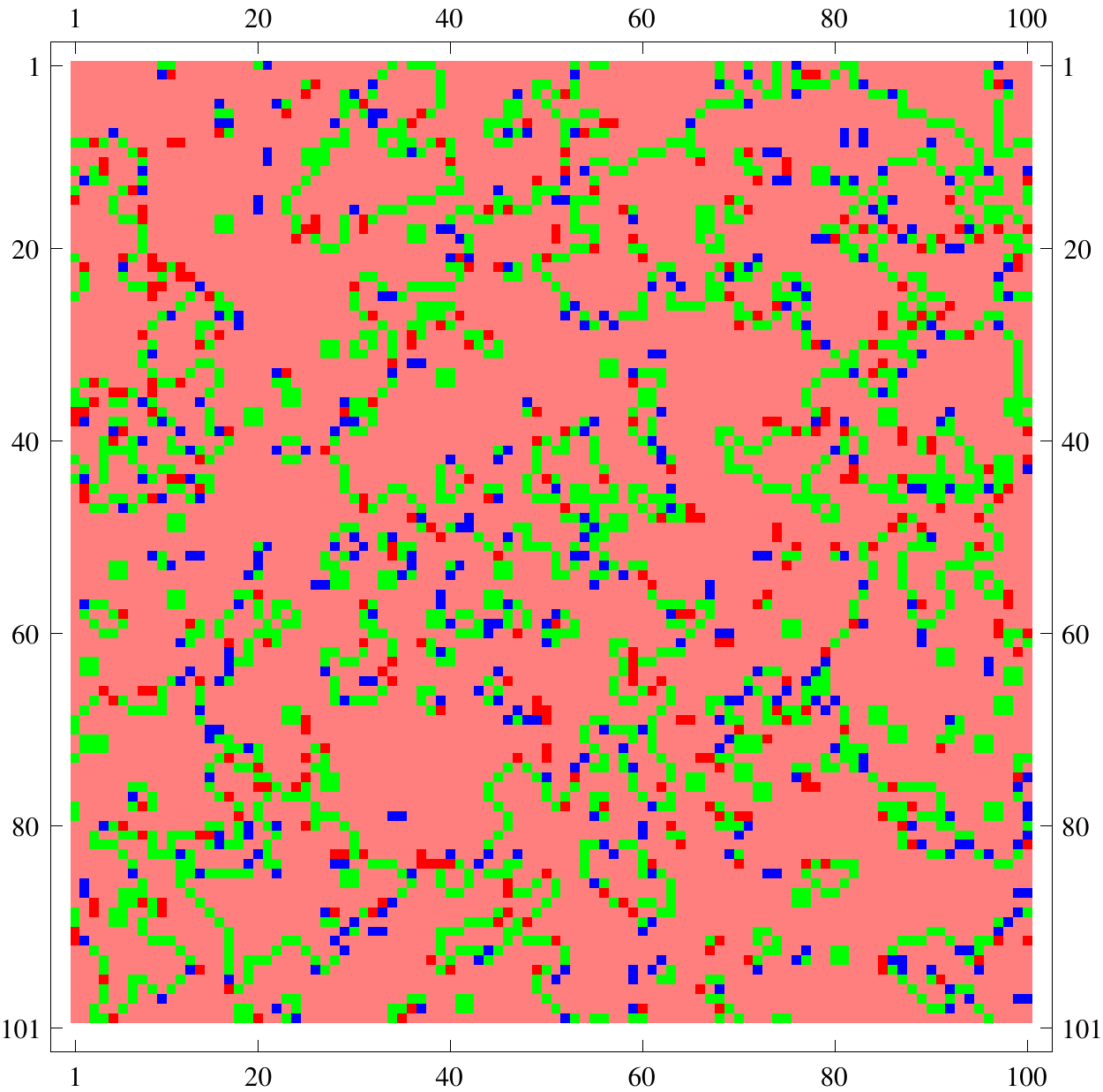}
\includegraphics[scale=0.22]{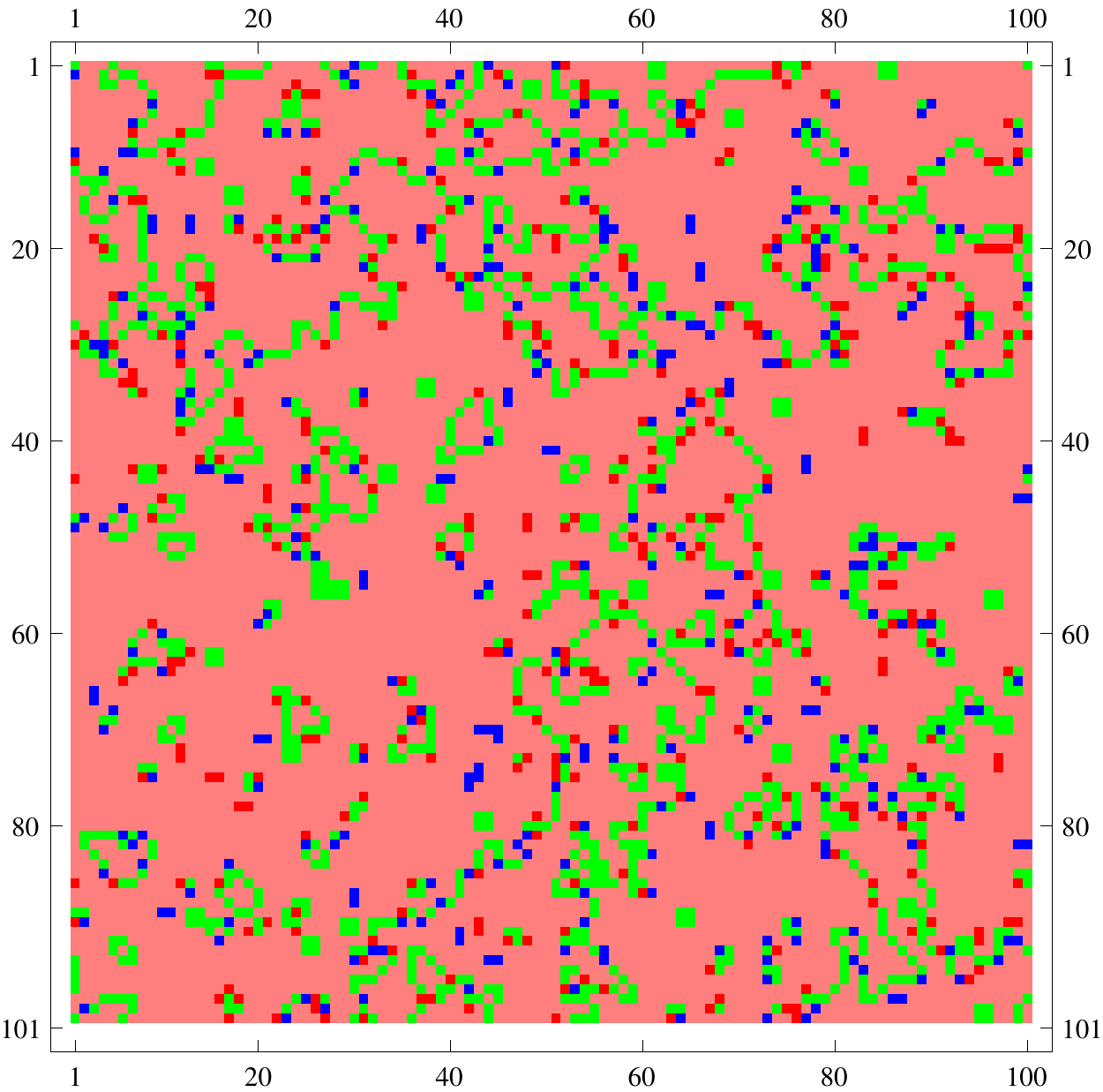}
\\
(a)
\vspace{0.3cm}
\\
\includegraphics[scale=0.22]{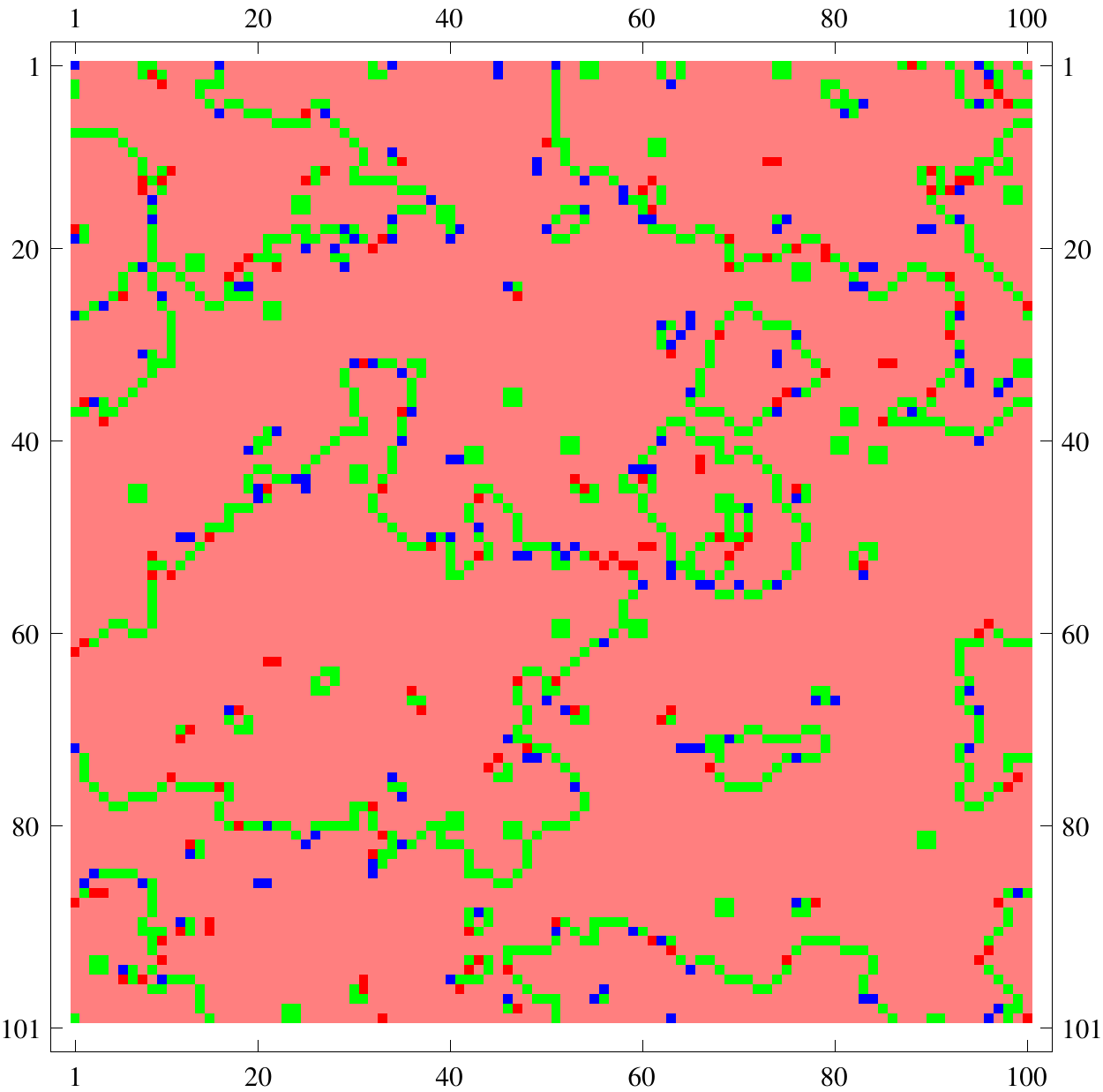}
\includegraphics[scale=0.22]{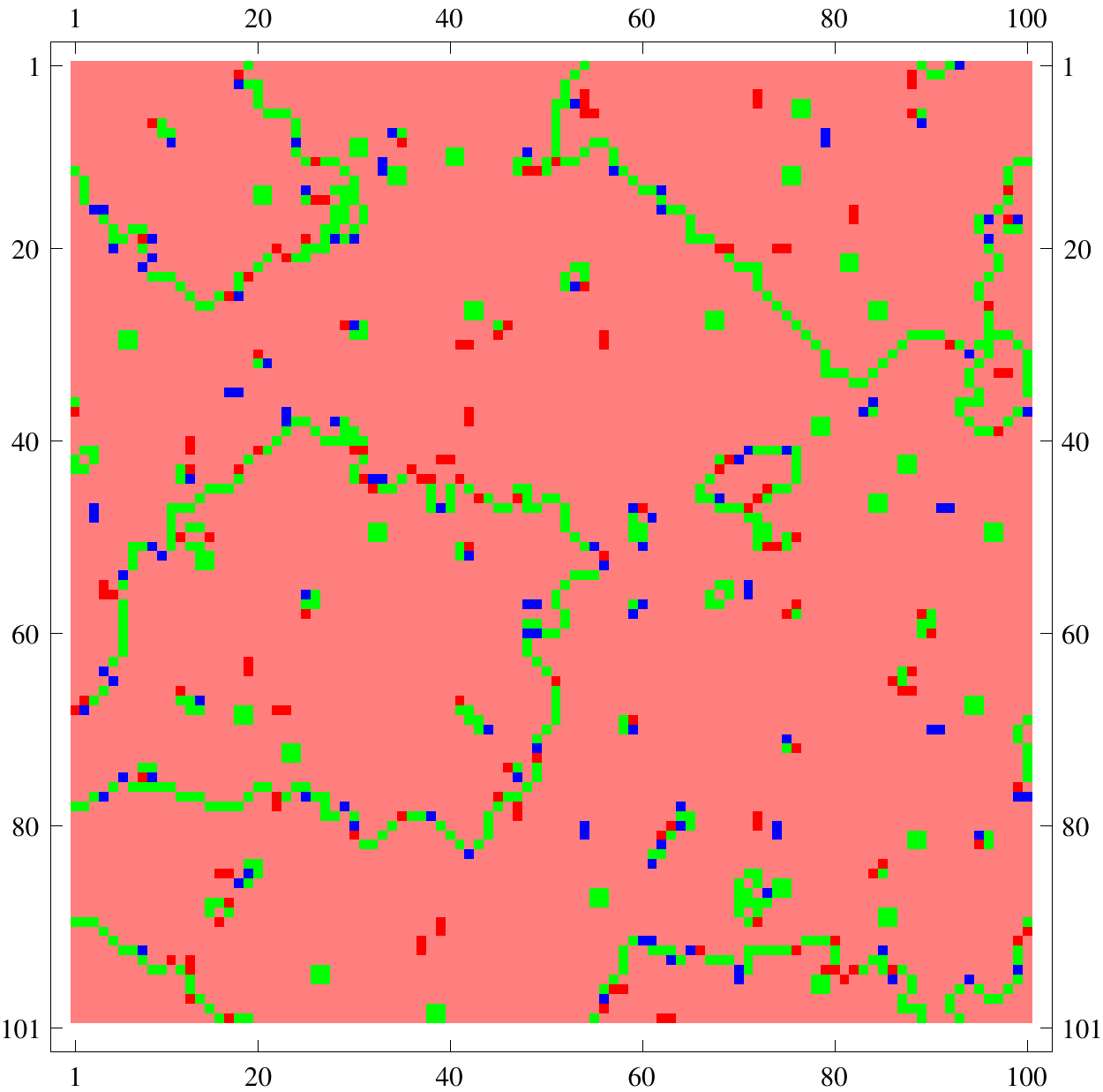}
\includegraphics[scale=0.22]{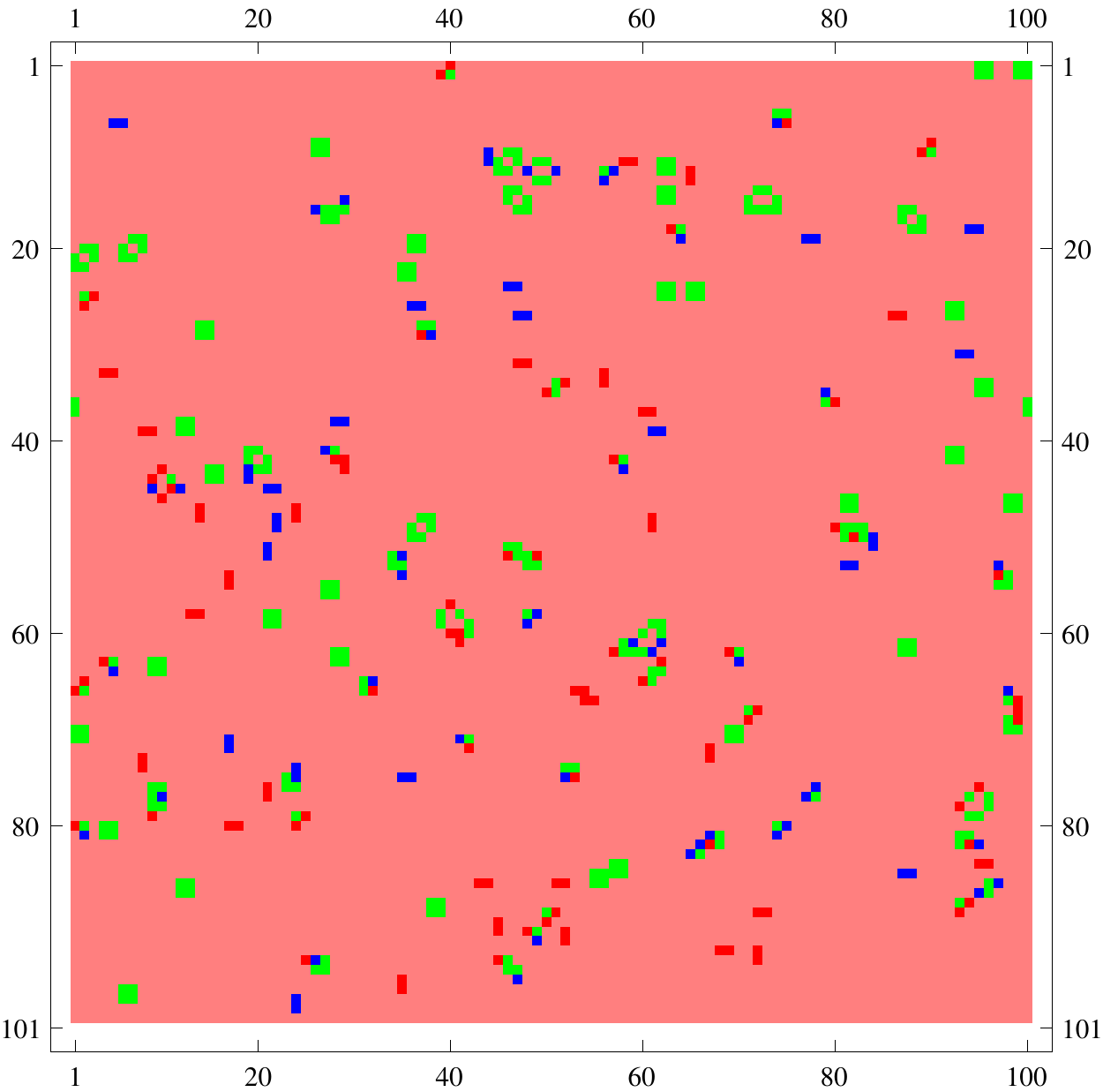}
\\
(b)
\end{centering}
\caption{(Color online.) Snapshots of  $L=10^2$ samples at $t\simeq 10^2, 10^3, 10^4$ MCs
after a quench from $\beta= 0$ to the critical point, $\beta_c=1.2$~(a) and into the AF phase, $\beta=1.36$~(b). 
Pink regions are $c$-AF ordered, 
green points correspond to $a,b$-FM vertices, 
red and blue points correspond to oppositely charged defects of type $e$. 
}
\label{snapshots}
\end{figure}

\paragraph{Correlation functions.}
Stochastic thermal evolution and canonical equilibration not only yield the vertex densities but also 
the correlation functions.
Figure~\ref{correlation-function}(a) shows the space-dependence of the two-point function 
$C(r,t)=L^{-2}\langle \sum_{i,j} S_{i,j}(t) S_{i+r,j+r}(t)\rangle$ at different times after a quench from $\beta = 0$ to 
$\beta_c= 1.2$ (log-log scale). $S_{i,j}$ denote the spins sitting along the edges of the $L\times L$ 
square lattice, with 
$S_{i,j}=+1$ if the spin points right or up and $S_{i,j}=-1$ otherwise.
$r$ is measured in units of $\ell/\sqrt{2}$. These data have been averaged over $10^3$ 
independent runs of a system with $L=60$.
At large times the curves approach the equilibrium asymptotic law characterized by 
$C(r,t\to\infty) \sim r^{-\eta}$ at the critical point, with an exponential cut-off.
As shown in the figure (black dotted line), the exponent $\eta$ remains 
equal to $1/4$ in the sixteen-vertex model 
as it was argued in~\cite{CavityMC-short}. This value coincides with 
the exact result of the eight-vertex~\cite{LiebWuBook, BaxterBook} and $2D$ Ising models.
Note that the numerical data show clear signs of finite size effects when $r\approx L$. 

In Fig.~\ref{correlation-function}(b) we show the behavior of $C(r,t)$ after a quench into the $c$-AF phase ($\beta=1.36$).  
As shown in~\cite{Levis2012, Budrikis2012-short}, 
the approach to equilibrium  
is fast if the initial state is a $T=0$ ground state whereas it is very slow and occurs 
{\it via} a coarsening process if the initial condition is a disordered high temperature one, as for the 
curves shown in the figure.  
In equilibrium correlations decay exponentially as $C(r,t\to\infty) \sim A \exp (-r/\xi) + M_-^2$ (dotted black line).
The asymptotic value $M_-^2\approx 0.73$ is 
consistent with 
the equilibrium staggered magnetization shown in Fig.~\ref{FiniteSize}, and the correlation length is $\xi(\beta=1.36) \approx 3$.
On the other hand, out-of-equilibrium spatial correlations decay to zero at large distances. 
The data for $C(r,t)$  at different times shown in Fig.~\ref{correlation-function}(b)  collapse onto a single curve 
when the length variable $r$ is rescaled by $t^{1/2}$ (see the inset). This scaling is accurate in a certain time window (i.e. the 
\emph{coarsening regime}): it fails at times larger than $\approx 2600$~MCs (as shown in the inset) and smaller than $\approx 100$~MCs. 
The snapshots in Fig.~\ref{snapshots} and the $t^{1/2}$ scaling strongly suggests that the system follows a curvature driven type of 
dynamics during the coarsening regime~\cite{Bray1994}.

These results imply that the samples obtained by using the rotating field 
protocol~\cite{Nisoli2010a-short}, where no correlations beyond first-neighbors were observed, have not achieved
equilibrium. On the contrary, the as-grown samples in~\cite{Morgan2011-short,Morgan2013} are likely to be near equilibrium.
However, as also shown by our numerical results, close to the phase transition 
critical slowing down sets in, and
in the whole AF phase slow coarsening dynamics emerge. 
It remains therefore to be understood
whether critical and subcritical samples have  {\it fully}
equilibrated. To settle this issue one should grow samples with a slower deposition rate
and measure, if possible, time-dependent observables {\it during} growth (such as the staggered magnetization
and two-times correlation functions). Another possibility is to analyze the spatial correlations.
Indeed, the facility of imaging the microstates both numerically and experimentally makes such study very appealing.
As an example, in Fig.~\ref{snapshots} we show snapshots of microscopic configurations of $L=10^2$ systems
after quenches from $\beta= 0$ to $\beta_c = 1.2$~(a) and $\beta=1.36$~(b). The first two panels in each row are 
out-of-equilibrium while the last ones show typical equilibrium configurations at the critical point and inside the AF phase
respectively, cfr. Fig.~\ref{correlation-function}. 
Large domains of ground state AF order form,
separated by domain walls made by $a$ and $b$ vertices. The size of the AF domains increases with time as
equilibrium is approached. These snapshots could be 
compared with MFM images of ASI configurations of~\cite{Morgan2011-short,Morgan2013} at the corresponding values of
$\beta$ (the experimental points closest to $\beta_c = 1.2$ and $\beta = 1.36$ correspond to 
the samples produced using a Ti underlayer with lattice constants $\ell = 466 \, \textrm{nm}$ and $\ell = 433 \, \textrm{nm}$,
respectively).

\paragraph{Conclusion.} 
In this letter we studied the equilibrium and out-of-equilibrium properties
of the $2D$ sixteen-vertex model using MC simulations and a sophisticate
cluster variational Bethe-Peierls approach, and we compared our results to 
the recent experimental data on ASI~\cite{Nisoli2010a-short,Morgan2011-short,Morgan2013}. 
We showed that the model describes
with very good accuracy the behavior of the densities of the different vertex types
of the ASI samples obtained by gradual deposition of
magnetic material on a square pattern 
as the lattice constant and the underlayer disorder are changed, resulting in a change of the
statistical weights of the different vertex types.  
This implies that the experimental samples of~\cite{Morgan2011-short,Morgan2013} are at---or at least
very close to---thermal equilibrium. It is important to point out that our interpretation does not
require any fitting parameter as the effective temperature introduced in~\cite{Nisoli2010a-short,Morgan2013}. 
We reveal the presence of a second order phase transition from a conventional high temperature
(large lattice constant, strong disorder) PM phase to a low temperature (small lattice constant, weak disorder)
staggered AF phase.
Such a phase transition could have a major impact on thermalization and full equilibration, and
could be unveiled  by measuring long-range spatial correlations in the experiments.

We close by insisting upon the fact that, although vertex models avoid all the complications
of (long-range) dipolar interactions, they provide a very good schematic framework
to study ASI from a theoretic perspective. The excitement around these samples as well 
as the intriguing excitation properties of spin-ice (emergence of magnetic 
monopoles and attached Dirac strings~\cite{Castelnovo2008a-short}) 
should encourage their study from a novel and more phenomenological perspective.

\paragraph{Acknowledgments:}
We thank  T. Blanchard, C. Castelnovo, C. Nisoli for very useful discussion
and G. Brunell, J. Morgan and C. Morrows  for lending their experimental data to us.
We acknowledge financial support from ANR-BLAN-0346 (FAMOUS).
 
\bibliographystyle{apsrev}
\bibliography{ASI-short}{}

\end{document}